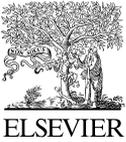

Available online at www.sciencedirect.com

ScienceDirect

Nuclear Physics B 993 (2023) 116281

www.elsevier.com/locate/nuclphysbQuantum Field Theory and Statistical Systems

# Gauge theory on fiber bundle of hypercomplex algebras

Hun Jang [a,b]

[a] *Center for Cosmology and Particle Physics, Department of Physics, New York University, 726 Broadway, New York, NY 10003, USA*
[b] *Institute for Theoretical Physics, and Riemann Center for Geometry and Physics, Leibniz University Hannover, Appelstraße 2, 30167 Hannover, Germany*Received 18 May 2023; received in revised form 13 June 2023; accepted 21 June 2023
Available online 22 June 2023
Editor: Hubert Saleur**Abstract**

I introduce a way of constructing a fiber bundle whose fibers are given by hypercomplex algebras and woven by appropriate structure group, and present that a novel gauge theory can be built on the hypercomplex fiber bundle. In this work, I aim to answer a question about how nature selects one preferred vacuum among degenerate physical vacua, called *vacuum selection problem*. In the end, I found presence of the impenetrable domain wall that prohibits phase transition between the two vacua. To be specific, I found that in this theory, one particular vacuum between two degenerate physical vacua for Higgs-like scalar potential can be dynamically chosen with priority due to intrinsic even parity of both a scalar field and its vacuum under a $\mathbb{Z}_2$ symmetry, even though its scalar potential is given to be $\mathbb{Z}_2$-symmetric under both odd- and even-parity transformations of the scalar field. This means that the vacuum selection problem can be resolved in this gauge theory. I suggest that this work may be a gateway to addressing the theoretical origin of the true physical vacuum that nature takes.© 2023 The Author(s). Published by Elsevier B.V. This is an open access article under the CC BY license (http://creativecommons.org/licenses/by/4.0/). Funded by SCOAP³.**Contents**

1. Introduction . . . . . . . . . . . . . . . . . . . . . . . . . . . . . . . . . . . . . . . . . 2
2. Hypercomplex algebra . . . . . . . . . . . . . . . . . . . . . . . . . . . . . . . . . . 3
3. Gauge theory on fiber bundle of hypercomplex algebras . . . . . . . . . . . . . . . 4
    3.1.    Brief review on non-trivial U(1) complex line bundle . . . . . . . . . . . . . 4*E-mail address:* hun.jang@nyu.edu.https://doi.org/10.1016/j.nuclphysb.2023.1162810550-3213/© 2023 The Author(s). Published by Elsevier B.V. This is an open access article under the CC BY license (http://creativecommons.org/licenses/by/4.0/). Funded by SCOAP³.

## 1. Introduction

It is enigmatic how nature selects one specific vacuum with priority among physical vacua of scalar potential. In Brout-Englert-Higgs (BEH) mechanism [1], after choosing the unitary gauge, Higgs scalar potential has two degenerate physical vacua. It is then required to choose a particular vacuum orientation between positive and negative directions of the Higgs field. In the end, the Higgs vacuum is determined merely by our choice during the spontaneous symmetry breaking. Accordingly, there is no any physical justification of such choice between the two physical vacua. Moreover, this situation equally arises for any theories considering spontaneous symmetry breaking, like supersymmetric and superstring theories. In this respect, I cast an open question: *How does nature make selection of one particular vacuum with priority among obtainable vacua?* I shall call this issue *vacuum selection problem*. Throughout this paper, I aim to answer this.

In the meantime, development of algebraic structures in mathematics has brought physicists many insights to find the laws of nature. We know that the *real algebra* appears everywhere in physics. The *complex algebra* has incredibly broadened our perspective on microscopic nature by enabling us to do quantum mechanics. The *grassmann algebra* enables us to describe fermion with spinor, which is the key ingredient of supersymmetric theories. In addition, various trials that utilize *quaternionic* [2] and *split-complex or hyperbolic algebras* [3] for physics have been studied in the recent decades. As interesting results recently discovered, it has been proposed that the three generations of the standard model (SM) may be realized by the algebra "$\mathbb{C} \otimes \mathbb{O}$" of complex and *octonionic* numbers [4]. Moreover, it has been suggested that the algebra "$\mathbb{R} \otimes \mathbb{C} \otimes \mathbb{H} \otimes \mathbb{O}$" including *real, complex, quaternionic, and octonionic* ones may be treated as a gateway to explaining some features of the SM like symmetry breaking [5] and one generation of SM Weyl representations [6]. In Ref. [7], the author seeks out reformulation of the SM by taking advantage of algebraic structure. Conclusively, these stories of using algebraic structures imply that *what physical information we can excavate depends on what mathematical language we use for physical reasoning*. In this sense, it is worthwhile to research new mathematical notion and its possible application to physics because there is no guarantee that the conventional approaches for physics would always be successful in clarifying physical principles in nature. What is the next? I note a so-called *hypercomplex number*, which was studied by Anthony Harkin and Joseph Harkin in 2004 [8]. To the best of my knowledge, I have never seen any application of the hypercomplex number to physics. In this paper, I take the first step of such application by considering the gauge principle [9] in Sec. 3.

What is the hypercomplex number? First of all, to better understand its nature, let us go over three familiar types of number. It is common that for a variable $x$ in $\mathbb{R}$, it is not possible to find a consistent solution to the equation $x^2 = -1$ that belongs to the set of real numbers $\mathbb{R}$. The imaginary unit $i$ such that $i^2 = -1$ is thus introduced to solve the equation, extending the set of real numbers $\mathbb{R}$ into a set of the so-called *complex numbers* $\mathbb{C}$. Next, consider an equation $x^2 = 1$ where the solution is given by $x = \pm 1$ when $x \in \mathbb{R}$. Here there is another example when $x \notin \mathbb{R}$, which is the *hyperbolic (or split-complex)* number whose imaginary unit is given by $j$ such that $j^2 = 1$. The last case is *Study (or Grassmann)* number $\lambda$ such that $\lambda^2 = 0$, which also defines its own independent plane.

Looking the previous numbers, one may ask a question. Can there exist each complex-like field $\mathbb{C}_p$ whose defining imaginary unit is given by $i_p$ such that $i_p^2 = p$ for each real number $p$? If so, we are able to have infinitely-many complex-like fields since cardinality of real field $\mathbb{R}$ is infinite. Motivated by the above question, the hypercomplex number has been established by the authors of Ref. [8], where they have found its algebra and geometry in a mathematically rigorous manner. In this work, I raise another questions about the hypercomplex algebra: *Can we weave such independent vector spaces consisting of hypercomplex algebras into a fiber bundle by introducing appropriate structure group like the usual case of tangent bundle? What kind of gauge theory can we obtain from such fiber bundle? What type of dynamics can the theory describe?* Throughout this paper, I intend to answer the first two of the above questions in Sec. 3, and the last in Sec. 4. In addition, I would like to point out that the idea of an (octonionic) imaginary unit that varies from point to point on a manifold was proposed in Ref. [10]. It would thus be interesting to study how to fit the idea of Ref [10] with the fiber-bundle scheme that this paper dealing with the imaginary unit varying over a smooth manifold proposes for the first time. This paper is organized as follows. In Sec. 2, I briefly introduce a definition of hypercomplex algebra, and its properties found in Ref. [8]. In Sec. 3, I present how to comprise a fiber bundle of the hypercomplex algebras, and how to build gauge theory on the fiber bundle. In Sec. 4, I study scalar field dynamics that the gauge theory predicts, and utilize this for resolving the vacuum selection problem mentioned above.

## 2. Hypercomplex algebra

In this section, I give a brief review of the *hypercomplex number* introduced in Ref. [8]. The hypercomplex number can be considered as a generalization of the complex number in $\mathbb{C}$. The idea of hypercomplex number starts with a nontrivial solution of the following algebraic equation

$$x^2 = p \in \mathbb{R}, \tag{2.1}$$

where $x$ is a variable. Two trivial solutions to Eq. (2.1) can be obtained if $x \in \mathbb{R}$ for positive values of $p$, and $x \in \mathbb{C}$ for negative values of $p$. By the way, someone may cast a question about whether there exist independent spaces spanned by each of the solutions $x$'s which do not belong to either $\mathbb{R}$ or $\mathbb{C}$ anymore. In Ref. [8], the authors confirm that such spaces can be constructed, which is called hypercomplex planes. They define the new planes in the way

$$\mathbb{C}_p \equiv \{ x + i_p y \mid x, y \in \mathbb{R}, \ i_p^2 = p \}, \tag{2.2}$$

whose possible operation is given by a conjugate operation $\star$ such that

$$i_p^\star \equiv -i_p. \tag{2.3}$$

The conjugate of a hypercomplex number $z$ is then given by

$$z^\star = x - i_p y. \tag{2.4}$$

For two hypercomplex numbers $z_1 \equiv x_1 + i_p y_1$ and $z_2 \equiv x_2 + i_p y_2$, the multiplication is defined by

$$z_1 z_2 = (x_1 + i_p y_1)(x_2 + i_p y_2) = x_1 x_2 + i_p(y_1 x_2 + x_1 y_2) + i_p^2 y_1 y_2$$
$$= (x_1 x_2 + p y_1 y_2) + i_p(y_1 x_2 + x_1 y_2). \tag{2.5}$$

In hypercomplex algebras, the corresponding norm is defined in the way

$$\|z\|_p \equiv \sqrt{z_p^\star z_p} = \sqrt{(x - i_p y)(x + i_p y)} = \sqrt{|x^2 - p y^2|}. \tag{2.6}$$

In addition, the authors of Ref. [8] also found that a relevant trigonometry can be given as follows:

$$\cosp(\theta) \equiv \begin{cases} \cos(\theta\sqrt{|p|}) & p < 0 \\ 1 & p = 0 \\ \cosh(\theta\sqrt{p}) & p > 0 \end{cases}, \quad \sinp(\theta) \equiv \begin{cases} \frac{1}{\sqrt{|p|}}\sin(\theta\sqrt{|p|}) & p < 0 \\ \theta & p = 0 \\ \frac{1}{\sqrt{p}}\sinh(\theta\sqrt{p}) & p > 0 \end{cases},$$

$$\tanp(\theta) \equiv \frac{\sinp(\theta)}{\cosp(\theta)}. \tag{2.7}$$

This leads to the following relation for the hypercomplex algebras and geometry, which is analogous to the usual Euler-formula ($e^{i\theta} = \cos\theta + i\sin\theta$)

$$e^{i_p \theta} = \cosp(\theta) + i_p \sinp(\theta). \tag{2.8}$$

Using this, it is possible to rewrite the hypercomplex number $z$ in terms of analogous radial and angular coordinates $(r_p, \theta)$ as follows

$$z = x + i_p y = r_p e^{i_p \theta}, \qquad r_p = \|z\|_p. \tag{2.9}$$

## 3. Gauge theory on fiber bundle of hypercomplex algebras

In this section, following the convention of Ref. [11], I begin with a concise review on a nontrivial U(1) complex line bundle on which the conventional U(1) gauge theory like electrodynamics is defined. Next, I propose a so-called *local hypercomplex algebra* which is the hypercomplex algebra whose imaginary unit depends on coordinates of the base manifold $M$. Then, by adopting this local hypercomplex algebra as a fiber, I show how to weave the fibers that are identified with local hypercomplex algebras into a proper fiber bundle through appropriate structure group, providing a novel gauge theory on the bundle.

### 3.1. Brief review on non-trivial U(1) complex line bundle

A local section $s$ of a complex line bundle $L \xrightarrow{\pi} M$ whose fiber at $x \in U \subset M$, i.e. $F_x$, is isomorphic to a trivial product of a singleton $\{x\} \subset U$ and one-dimensional complex field $\mathbb{C}$, i.e. $F_x \approx \{x\} \times \mathbb{C}$:

$$s(x) = z(x) e(x) \in L, \tag{3.1}$$

where $z(x) = a(x) + ib(x) \in \mathbb{C}$ for real numbers $a, b \in \mathbb{R}$ and the imaginary unit $i$ such that $i^2 = -1$, and $e(x)$ is a basis of the local section $s(x)$.

Equivalently, such complex line bundle can be represented in terms of two real line bundles. We observe that the local section can be rewritten as

$$s(x) = z(x)e(x) = a(x)e(x) + b(x)ie(x) \equiv a(x)e_R(x) + b(x)e_I(x), \tag{3.2}$$

where we define $(e_R(x) = e(x), e_I(x) = ie(x))$, which is the basis of the bundle $L \xrightarrow{\pi} M$ whose fiber at $x \in U \subset M$ is now identified with $F_x \approx \{x\} \times \mathbb{R}^2$.

The *connection or covariant exterior derivative* $\nabla$ of the complex line bundle $L \xrightarrow{\pi} M$ is defined by a map

$$\nabla : \Omega^0(M, L) \longrightarrow \Omega^1(M, L) \tag{3.3}$$

where a space of k-forms with values in $L$ is defined by $\Omega^k(M, L) \equiv \mathcal{C}^\infty(M, L \otimes \wedge^k T^*M)$ for any integer $k \in \mathbb{N}$. Then, in general, we find the covariant exterior derivative of a local section $s(x)$, which is given by

$$\nabla s(x) = \nabla(z(x)e(x)) = dz(x) \otimes e(x) + z(x)\nabla e(x), \tag{3.4}$$

where $dz(x) = \partial_\mu z(x) dx^\mu$. If $e(x)$ is given by the local basis of the line bundle whose structure group is nontrivial, its covariant exterior derivative is given by

$$\nabla e(x) = iA(x)e(x) \in \Omega^1(M, F_x), \tag{3.5}$$

where we introduce imaginary unit $i$ since we assume Hermitian metric for giving a real scalar value, and one form $A(x) = A_\mu dx^\mu$ with values in $\mathbb{R}$ in which $A_\mu$ is called *gauge field*. Moreover, if $e(x) = 1 \in F_x$, then we have

$$\nabla 1 = iA(x) \otimes 1 \in \Omega^1(M, F_x). \tag{3.6}$$

We thus obtain the covariant exterior derivative of a local section $s(x)$ on the line bundle in the way

$$\nabla s(x) = \left[dz(x) + iA(x)z(x)\right]e(x) = \left[\left(\partial_\mu z(x) + iA_\mu(x)z(x)\right)dx^\mu\right]e(x)$$
$$\equiv \left[D_\mu z(x) dx^\mu\right]e(x) \equiv \left[Dz(x)\right]e(x) \in \Omega^1(M, F_x), \tag{3.7}$$

where $D_\mu$ is the covariant derivative.

### 3.2. Local hypercomplex algebras as trivial hypercomplex line bundle

**Local hypercomplex algebras.** Let us start with a hypercomplex algebra $\mathbb{C}_p$ for some $p$, which was given by

$$\mathbb{C}_p \equiv \{ z = a + i_p b \mid \forall a, b \in \mathbb{R}, \; i_p^2 = p \in \mathbb{R} \}. \tag{3.8}$$

Now I propose an idea that the parameter $p$ is not a constant but a smooth function over some manifold $M$. We may then imagine that for $x \in M$, and a function $p : M \to \mathbb{R}$

$$\mathbb{C}_{p(x)} \equiv \{ z = a + i_{p(x)} b \mid \forall a, b \in \mathbb{R}, \; \forall x \in M : i_p^2 = p(x) \in \mathbb{R} \}. \tag{3.9}$$

At first glance, one may question how we can think of a derivative of the local generalized-complex imaginary unit $i_{p(x)}$ with respect to the continuous variable $x \in M$. We can answer this by recognizing the local unit as a basis of vector space and taking advantage of fiber bundle structure. First, we should be careful about what the local imaginary unit $i_p$ defines. Each unit at a different point over the manifold spans its own independent vector space. This implies that it is very similar to the case of "tangent bundle." Here, we face the problem about how to compare two vectors from different tangent spaces arises, which subsequently gives rise to the issue of proper differentiation. We know that the issue can easily be washed out through the concept of connection and covariant derivative from fiber bundle structure. In this respect, we set up a strategy for constructing a fiber bundle using the hypercomplex algebra.

I propose to comprise a *trivial* fiber bundle as a straightforward starting point. We consider a fiber bundle $V \xrightarrow{\pi} M$ whose total, base spaces, and projection map are given by a vector space $V$, a smooth manifold $M$, and a surjective map $\pi$ respectively. The fiber at some point $x \in M$ is then defined by $V_x \equiv \pi^{-1}\{x\}$. The element of the fiber space $V_x$ is now given by a section $s(x) = s_i(x)e_i(x) \in V_x$ where $s_i(x), e_i(x)$ are components and basis of the section. The structure group $G$ of the fiber bundle will be chosen as our flavor.

Since we are interested in making a trivial fiber bundle, the corresponding structure group should be a trivial group $G = \{e\}$ whose members are given solely by the identity element "$e$." Then, let us begin with a product space $M \times \mathbb{R}^2$ for some manifold $M$ and real field $\mathbb{R}$, so that the typical fiber space is given by $F = \mathbb{R}^2$. Suppose that there exists an isomorphism $h$ from the product space to some vector space $V$ defined by

$$h : M \times \mathbb{R}^2 \longrightarrow V \tag{3.10}$$

$$(x, a, b) \mapsto v = h(x, a, b) \in V_x \subset V, \tag{3.11}$$

where $x \in M$, $a, b \in \mathbb{R}$, and $v \in V$. Eventually, we notice the equivalence $s_1 = a, s_2 = b$. Now we assume that the basis vectors are defined by

$$e_1(x) = 1, \ e_2(x) = i_{p(x)} \in V_x \subset V, \tag{3.12}$$

such that

$$e_1(x)^2 = e_1(x) = 1 \in V_x, \tag{3.13}$$

$$e_1(x)e_2(x) = e_2(x)e_1(x) = e_2(x) = i_{p(x)} \in V_x, \tag{3.14}$$

$$e_2(x)^2 = i_{p(x)}^2 = p(x) \cdot 1 \in V_x. \tag{3.15}$$

As a result, we identify the following

$$h(x, a, b) = v = s(x) = a(x) \cdot 1 + b(x) \cdot i_{p(x)} \in V_x, \tag{3.16}$$

and that our trivial vector bundle is specified as $(V \cong M \times \mathbb{R}^2 \xrightarrow{\pi} M, G = \{e\})$. Again, our structure group here is a trivial group $G = \{e\}$, meaning that we are able to have global sections over the base manifold and no gauge transformations of the relevant gauge field.

Next, one may question about the connection $\nabla$ of the trivial fiber bundle. Before going into the main story about it, I present here a summary about connection, covariant (exterior) derivative, and curvature. The *connection* $\nabla$ of a fiber bundle $L \xrightarrow{\pi} M$ is endowed as a map

$$\nabla : \Omega^0(M, L) \longrightarrow \Omega^1(M, L) \tag{3.17}$$

and the corresponding covariant exterior derivative $d_\nabla$ is defined as

$$d_\nabla : \Omega^k(M, L) \longrightarrow \Omega^{k+1}(M, L) \tag{3.18}$$

where a space of k-forms with values in $L$ is defined by $\Omega^k(M, L) \equiv \mathcal{C}^\infty(M, L \otimes \wedge^k T^*M)$ for any integer $k \in \mathbb{N}$, and $d_\nabla = \nabla$ when $k = 0$. The operator obeys the same Leibniz rule of an exterior derivative $d$ given by

$$d(\omega \wedge \eta) = d\omega \wedge \eta + (-1)^k \omega \wedge d\eta, \tag{3.19}$$

$$d_\nabla(\omega \wedge \eta) = d_\nabla \omega \wedge \eta + (-1)^k \omega \wedge d_\nabla \eta, \tag{3.20}$$

where $\omega \in \Omega^k(M, L)$ and $\eta \in \Omega^l(M, L)$ for any $k, l \in \mathbb{N}$. The interesting fact about the covariant exterior derivative is that for any section $s \in \Omega^0(M, L) = \Gamma(M, L)$ and two tangent vectors $X, Y \in TM$, we have

$$(d_\nabla^2 s)(X, Y) = R(X, Y)s \tag{3.21}$$

where $R$ is the Lie-algebra-valued curvature two-form.

Getting back to the story of the trivial vector bundle, we can consider a connection structure as follows. Specifically, from the algebra (3.13), we find that

$$\nabla e_1(x)^2 = 2e_1(x)\nabla e_1(x) = \nabla e_1(x) \implies (2e_1(x) - 1)\nabla e_1(x) = 0 \implies \nabla e_1(x) = \nabla 1 \tag{3.22}$$

From the algebra (3.15), we observe that

$$\nabla e_2(x)^2 = 2e_2(x)\nabla e_2(x) = \nabla p(x) = dp(x) \cdot 1 + p(x)\nabla 1 = dp(x) \cdot 1$$
$$\implies 2i_{p(x)}\nabla i_{p(x)} = dp(x) \cdot 1$$
$$\implies 2p(x)\nabla i_{p(x)} = dp(x) i_{p(x)}$$
$$\implies \nabla i_{p(x)} = \frac{dp(x)}{2p(x)} i_{p(x)} \in V_x \tag{3.23}$$

Changing some notations to

$$F_x = \pi^{-1}(\{x\}) = V_x \equiv \mathbb{C}_{p(x)}, \quad F = \mathbb{R}^2$$
$$V = \bigcup_{x \in M} \{x\} \times \mathbb{R}^2 = \bigcup_{x \in M} V_x = \bigcup_{x \in M} \mathbb{C}_{p(x)} \equiv \mathbb{C}_{p(M)} \cong M \times \mathbb{R}^2, \tag{3.24}$$

we summarize the following: a *trivial* vector bundle $(\mathbb{C}_{p(M)} \cong M \times \mathbb{R}^2 \xrightarrow{\pi} M, G = \{e\})$ such that

$$s(x) = a(x) \cdot 1 + b(x) \cdot i_{p(x)} \in \mathbb{C}_{p(x)} = \mathbb{C}_{p(M)}|_x$$
$$\nabla e_1(x) = \nabla 1 = 0, \quad \nabla e_2(x) = \nabla i_{p(x)} = \frac{dp(x)}{2p(x)} i_{p(x)} \in \mathbb{C}_{p(x)} = \mathbb{C}_{p(M)}|_x, \tag{3.25}$$

which gives rise to

$$\nabla s(x) = [da(x) \cdot 1 + a(x)\nabla 1] + [db(x) \cdot i_{p(x)} + b(x)\nabla i_{p(x)}]$$
$$= da(x) \cdot 1 + i_{p(x)}\left[db(x) + b(x)\frac{dp(x)}{2p(x)}\right]. \tag{3.26}$$

It turns out that the trivial vector bundle $\mathbb{C}_{p(M)}$ is itself a local algebra, defining a *local* hypercomplex field which can be considered as a new kind of "Field" in the mathematical sense.

**Duality between complex and hypercomplex algebras.** For $p(x) \equiv -f(x)$ for $f(x) > 0 \in \mathbb{R}$, we have $i_{-f(x)}$ such that $i_{-f(x)}^2 = -f(x)$. Now we introduce a particular combination

$$\hat{i}_{-f(x)} \equiv \frac{i_{-f(x)}}{\sqrt{f(x)}} \quad \text{such that} \quad \hat{i}_{-f(x)}^2 = -1 \in \mathbb{C}_{p(x) = -f(x)}. \tag{3.27}$$

Interestingly, this combination satisfies an interesting property, which is

$$\nabla \hat{i}_{-f(x)} = d\left(\frac{1}{\sqrt{f(x)}}\right) i_{-f} + \frac{1}{\sqrt{f(x)}} \nabla i_{-f} = -\frac{df(x)}{2f(x)\sqrt{f(x)}} i_{-f} + \frac{df(x)}{2f(x)} \frac{1}{\sqrt{f(x)}} i_{-f}$$
$$= 0, \tag{3.28}$$

which means that $\hat{i}_{-f(x)}$ is covariantly constant. Notice that if $f(x) \leq 0$, we are not able to possess such a good combination which can be covariantly constant on the fiber bundle. In fact, this implies that we are able to treat $\hat{i}_{-f}$ as an analogue of the imaginary unit $i$ of the complex algebra. I speculate that the complex and hypercomplex algebras can be dual to each other through the following isomorphism

$$i \longleftrightarrow \hat{i}_{-f(x)}. \tag{3.29}$$

However, it should be noticed that they are not exactly the same to each other, but spanning individual spaces. Considering such duality, one may define analogue of Pauli matrices, say $\hat{\sigma}_1, \hat{\sigma}_2, \hat{\sigma}_3$, over the hypercomplex algebras in the way that

$$\hat{\sigma}_1 = \begin{pmatrix} 0 & 1 \\ 1 & 0 \end{pmatrix}, \quad \hat{\sigma}_2 = \begin{pmatrix} 0 & -\hat{i}_{-f} \\ \hat{i}_{-f} & 0 \end{pmatrix}, \quad \hat{\sigma}_3 = \begin{pmatrix} 1 & 0 \\ 0 & -1 \end{pmatrix}, \tag{3.30}$$

which are also covariantly constant. With this analogue, we are able to handle spinor algebras for fermion by replacing the imaginary unit $i$ of the conventional Clifford algebras with the special hypercomplex imaginary unit $\hat{i}_{-f}$. Again, the two distinct units are not identical since they define different mathematical spaces. Furthermore, it is easy to check that for the combination $\hat{i}_{-f(x)}$ and some real function $\Lambda$, we are able to have the following formula

$$e^{\hat{i}_{-f(x)}\Lambda} = \cos(\Lambda) + \hat{i}_{-f(x)} \sin(\Lambda). \tag{3.31}$$

**Essential condition.** The condition given by

$$\forall x \in M : f(x) > 0 \in \mathbb{R} \tag{3.32}$$

is the most important requirement in this gauge theory. This enables us to take advantage of the duality, and define the Dirac delta function in the space $\mathbb{C}_{-f(x)}$

$$\delta(x - \alpha) \equiv \frac{1}{2\pi} \int_{-\infty}^{\infty} dp e^{\hat{i}_{-f(x)} p(x-\alpha)}, \tag{3.33}$$

which is one-dimensional case, and the plane wave solutions of the Klein-Gordon equation of motion for scalar field, i.e. $\nabla_\mu \nabla^\mu + m^2 = 0$,

$$\varphi(x) \equiv \int \frac{d^4k}{(2\pi)^4} [a(\vec{k}) e^{\hat{i}_{-f(x)} k \cdot x} + a^\star(\vec{k}) e^{-\hat{i}_{-f(x)} k \cdot x}], \tag{3.34}$$

where $a, a^\star$ should be considered as the ladder operators. Again, notice that the duality and results Eqs. (3.27), (3.33), (3.31), and (3.34) are valid only if $f(x)$ is positive-definite and nowhere vanishing over the base manifold $x \in M$. Furthermore, in Sec. 4, we will see that only under this condition, kinetic terms of some field degrees of freedom can possess right sign; that is, they can be ghost-free.

On the contrary, what would happen if $f(x) = -|f(x)| < 0$? This negative function leads us to define another covariantly-constant unit $\hat{i}_{|f(x)|} \equiv i_{|f(x)|}/\sqrt{|f(x)|}$ where $i^2_{|f(x)|} = |f(x)|$ in a consistent way, giving $\hat{i}^2_{|f(x)|} = 1$. In this case, as analogy of the case when $f(x) > 0$, we may

find that the unit $\hat{i}_{|f(x)|}$ has another type of duality that $\hat{i}_{|f(x)|} \leftrightarrow j$ where $j$ is defined by the hyperbolic unit such that $j^2 = 1$. In particular, the Euler-like formula corresponding to the new unit now consists of hyperbolic functions, i.e. $e^{\hat{i}_{|f(x)|}\Lambda} = \cosh(\Lambda) + \hat{i}_{|f(x)|}\sinh(\Lambda)$. Here we note that unlike the positive function case $f(x) > 0$, it is not able for us to obtain the (periodic) plane wave solutions (3.34) within the hypercomplex algebra of the unit $\hat{i}_{|f(x)|}$ because transforming between trigonometric and hyperbolic functions requires the imaginary unit that is equivalent to $i$ such that $i^2 = -1$ as in the case of the duality (3.29); for example, $\cosh ix = \cos x$ for some variable $x$ and the imaginary unit $i$ in the complex algebra. This means that such unavailability of the periodic functions in the algebra made of the negative function $f(x) < 0$ cannot reproduce the usual dynamics of field theory. After all, this failure justifies necessity of the condition (3.32).

### 3.3. Gauge theory from non-trivial hypercomplex line bundle

We are now in a position to develop non-trivial vector bundles whose structure group is not trivial.

**Principal G-bundle.** First of all, we explore a proper principal G-bundle that can be associated with the fiber bundle of hypercomplex algebras. Let $P$ and $\pi_P$ be a total space and a projection map, respectively. Then, we define a principal G-bundle

$$P \xrightarrow{\pi_P} M, \tag{3.35}$$

whose fiber at $x \in M$, i.e. $P_x$:

$$P_x \equiv G(x) = \{g(x) = e^{i_{p(x)} T \theta(x)} \in \mathbb{C}_{p(x)} | \, \theta(x) \in \mathbb{R}\}, \tag{3.36}$$

where $T \in \mathbb{R}$ is a generator of the gauge group with respect to the gauge parameter $\theta(x)$. Notice that over the local hypercomplex algebra $\mathbb{C}_{p(M)}$, we are able to define an independent Lie group, i.e. $P_x = G(x)$, at each point over the base manifold $M$. Then, we finally obtain the principal G-bundle:

$$G(M) \equiv \bigcup_{x \in M} G(x). \tag{3.37}$$

**Associated G-bundle.** Next, we find a fiber bundle structure necessary to construct the relevant associated G-bundle. First, let us define a non-trivial hypercomplex frame bundle $E \xrightarrow{\pi'} M$ whose fiber at $x \in M$ is given by $E_x = {\pi'}^{-1}(\{x\}) = \mathbb{R} \otimes \mathbb{C}_{p(x)} = \{v | v = ce(x)\}$ for some $c(x) \in \mathbb{R}$. Its basis vector $e(x) \in E_x$ is one-dimensional on the hypercomplex space and can transform under a non-trivial structure group $g(x) \in G(x) \subset G(M)$. Next, taking this as a structure group, we build a non-trivial tensor-product bundle as *hypercomplex line bundle* defined by

$$L \equiv \mathbb{C}_{p(M)} \otimes E \xrightarrow{\pi \otimes \pi'} M \tag{3.38}$$

whose fiber at $x \in M$ is given by $L_x = (\pi \otimes \pi')^{-1}(\{x\}) = \mathbb{C}_{p(x)} \otimes E_x$. The element of the fiber at $x \in M$ is given by a local section $s(x)$ of $L$ (or so-called "hypercomplex-valued scalar field") as follows

$$\begin{aligned} s(x) &= a(x)(1 \otimes e(x)) + b(x)(i_{p(x)} \otimes e(x)) = (a(x) + b(x)i_{p(x)})e(x) \in L_x \\ &= \mathbb{C}_{p(x)} \otimes E_x. \end{aligned} \tag{3.39}$$

**Gauge covariance.** Allowing (gauge) connection $\nabla$ to the non-trivial bundle $L$ as well, we are able to find its action onto the local section $s(x)$. This implies that the connection of the basis $e(x) \in E_x$ accompanies a gauge connection one-form $A(x)$ (or gauge field $A_\mu$, i.e. $A(x) \equiv TA_\mu(x)dx^\mu \in T^*M$), which has something to do with gauge redundancy. As an analogue of $U(1)$ group, we may define the covariant exterior derivative of the local basis in the way

$$\nabla e(x) = A(x) i_{p(x)} e(x) \in \Omega^1(M, L_x). \tag{3.40}$$

I point out that the local connection one-from (or gauge field) $A(x) = TA_\mu dx^\mu$ must be uniquely characterized by a choice of the smooth function $p(x)$ because this determines the nature of the fiber bundle as well as gauge transformations of hypercomplex-valued local section or scalar field and the gauge connection one-form (or gauge field). That is, if such smooth function is different, then the corresponding gauge field must be distinct.

The covariant exterior derivative of a local section or the scalar field is then given by

$$\begin{aligned}\nabla s(x) &= \nabla(z(x)e(x)) = \nabla z(x)e(x) + z(x)\nabla e(x) \\ &= [\nabla z(x) + i_{p(x)} A(x) z(x)] e(x) \equiv [Dz(x)]e(x)\end{aligned} \tag{3.41}$$

where we define the minimal coupling (i.e. covariant derivative)

$$\begin{aligned}D &\equiv \nabla + i_{p(x)} A(x) \implies dx^\mu D_\mu \equiv dx^\mu \nabla_\mu + i_{p(x)} TA_\mu(x) dx^\mu \\ &\implies D_\mu = \nabla_\mu + i_{p(x)} TA_\mu,\end{aligned} \tag{3.42}$$

which sends a hypercomplex number in $\mathbb{C}_{p(x)}$ to a hypercomplex one-form, or equivalently we get

$$\begin{aligned}\nabla s(x) &= \nabla\Big(a(x)e(x) + b(x) i_{p(x)} e(x)\Big) \\ &= da(x)e(x) + a(x)\nabla e(x) + db(x) i_{p(x)} e(x) + b(x)(\nabla i_{p(x)})e(x) \\ &\quad + b(x) i_{p(x)} (\nabla e(x)) \\ &= da(x)e(x) + a(x) A(x) i_{p(x)} e(x) + db(x) i_{p(x)} e(x) \\ &\quad + b(x)(d\ln\sqrt{|p(x)|} i_{p(x)}) e(x) + b(x) i_{p(x)} (A(x) i_{p(x)} e(x)) \\ &= \Big(da(x) + b(x) p(x) A(x)\Big) e(x) \\ &\quad + \Big(db(x) + a(x) A(x) + b(x) d\ln\sqrt{|p(x)|}\Big) i_{p(x)} e(x).\end{aligned} \tag{3.43}$$

**Gauge transformations.** When taking transformation of the basis $e(x)$ on the same fiber in the way

$$e(x) \longrightarrow e'(x) = g(x)e(x), \tag{3.44}$$

we observe that its connection result should also change in the way

$$\nabla e'(x) = A'(x) i_{p(x)} e'(x), \tag{3.45}$$

where $A'$, $e'$ are gauge-transformed. In fact, we can read gauge transformation of the gauge field by solving the above equation. We see that the left-hand side of this equation gives

$$\begin{aligned}
\nabla e'(x) &= \nabla(g(x)e(x)) = \nabla g(x) \cdot e(x) + g(x) \cdot \nabla e(x) \\
&= g(x)\nabla(i_{p(x)}\theta(x))e(x) + g(x)\nabla e(x) \\
&= g(x)\Big(\nabla i_{p(x)}\theta(x)e(x) + i_{p(x)}d\theta(x)e(x) + \nabla e(x)\Big) \\
&= g(x)\Big(\frac{dp(x)}{2p(x)}i_{p(x)}\theta(x)e(x) + i_{p(x)}d\theta(x)e(x) + A(x)i_{p(x)}e(x)\Big) \\
&= g(x)\Big(\frac{dp(x)}{2p(x)}\theta(x) + d\theta(x) + A(x)\Big)i_{p(x)}e(x),
\end{aligned} \quad (3.46)$$

while its right-hand side is given by

$$\nabla e'(x) = A'(x)i_{p(x)}e'(x) = A'(x)i_{p(x)}g(x)e(x). \quad (3.47)$$

By comparing these two, we observe that the gauge transformation of the gauge connection one-form is found to be

$$A'(x) = A(x) + d\theta(x) + (d\ln\sqrt{|p(x)|})\theta(x). \quad (3.48)$$

Equivalently, the gauge transformation of the gauge field is thus given by

$$A'_\mu(x) = A_\mu(x) + \partial_\mu\theta(x) + (\partial_\mu\ln\sqrt{|p(x)|})\theta(x). \quad (3.49)$$

The current form of this transformation looks very different from the conventional one in the U(1) gauge theory like electrodynamics.

**Curvature and field strength.** Let us compute the curvature two-from. Using the curvature identity (3.21), we find

$$\begin{aligned}
d^2_\nabla e(x) &= d_\nabla(A(x)i_{p(x)}e(x)) = dA(x)i_{p(x)}e(x) - A(x) \wedge \nabla(i_{p(x)}e(x)) \\
&= dA(x)i_{p(x)}e(x) - A(x) \wedge \Big(\nabla i_{p(x)}e(x) + i_{p(x)}\nabla e(x)\Big) \\
&= dA(x)i_{p(x)}e(x) - A(x) \wedge \Big(d\ln\sqrt{|p(x)|}i_{p(x)}e(x) + i^2_{p(x)}A(x)e(x)\Big) \\
&= dA(x)i_{p(x)}e(x) - A(x) \wedge \Big(d\ln\sqrt{|p(x)|}i_{p(x)}e(x) + p(x)A(x)e(x)\Big) \\
&= \Big(dA(x) - A(x) \wedge d\ln\sqrt{|p(x)|}\Big)i_{p(x)}e(x) = R(x)e(x) \\
&\implies R(x) = \Big(dA(x) - A(x) \wedge d\ln\sqrt{|p(x)|}\Big)i_{p(x)} \equiv F(x)i_{p(x)},
\end{aligned} \quad (3.50)$$

where we used $A(x) \wedge A(x) = A_{(\mu}A_{\nu)}dx^{[\mu} \wedge dx^{\nu]} = 0$ and $i^2_{p(x)} = p(x)$, and define a field strength two-form

$$F(x) \equiv dA(x) - A(x) \wedge d\ln\sqrt{|p(x)|} = \frac{1}{2!}F_{\mu\nu}dx^\mu \wedge dx^\nu, \quad (3.51)$$

where

$$F_{\mu\nu} \equiv \partial_\mu A_\nu - \partial_\nu A_\mu - (A_\mu\partial_\nu\ln\sqrt{|p|} - A_\nu\partial_\mu\ln\sqrt{|p|}). \quad (3.52)$$

Notice that this two-form is gauge invariant. Using Eqs. (3.50) and (3.52), we are able to construct a gauge-invariant kinetic action of the gauge field as follows

$$\mathcal{L} = -\frac{1}{4g^2} R^\star_{\mu\nu} R^{\mu\nu} = \frac{p}{4g^2} F_{\mu\nu} F^{\mu\nu}$$
$$= \frac{p}{4g^2} \left| \partial_\mu A_\nu - \partial_\nu A_\mu - (A_\mu \partial_\nu \ln \sqrt{|p|} - A_\nu \partial_\mu \ln \sqrt{|p|}) \right|^2 \quad (3.53)$$

where $g$ is a gauge coupling constant, and the mass term of the gauge field is absent because it is not gauge invariant. Of course, it is possible to include such mass term once taking Stueckelberg trick [12]. We observe that $p$ must be negative-definite for the gauge field to be ghost-free. Hence, if we define $p(x) \equiv -f(x)$ for some real function $f(x) > 0 \in \mathbb{R}$, the action is rewritten as

$$\mathcal{L} = -\frac{f}{4g^2} F_{\mu\nu} F^{\mu\nu} = -\frac{f}{4g^2} \left| \partial_\mu A_\nu - \partial_\nu A_\mu - (A_\mu \partial_\nu \ln \sqrt{f} - A_\nu \partial_\mu \ln \sqrt{f}) \right|^2. \quad (3.54)$$

The gauge transformation of the field strength is given by

$$\begin{aligned} F' &= dA' - A' \wedge d\ln\sqrt{|p|} \\ &= dA + d^2\theta + d^2\ln\sqrt{|p|}\theta - d\ln\sqrt{|p|} \wedge d\theta - (A + d\theta + (d\ln\sqrt{|p|})\theta) \wedge d\ln\sqrt{|p|} \\ &= dA + 0 + 0 - d\ln\sqrt{|p|} \wedge d\theta - A \wedge d\ln\sqrt{|p|} - d\theta \wedge d\ln\sqrt{|p|} + 0 \\ &= F - d\ln\sqrt{|p|} \wedge d\theta + d\ln\sqrt{|p|} \wedge d\theta = F. \end{aligned} \quad (3.55)$$

Furthermore, we obtain the Bianchi identity from the equation $d_\nabla(Rs) = 0$ on any section $s \in L$. Taking $s = e(x)$ which is the basis vector of a local section, we see that

$$\begin{aligned} d_\nabla(Re) &= d_\nabla(F(x)i_{p(x)}e(x)) = dF(x)i_{p(x)}e(x) + F(x) \wedge \nabla(i_{p(x)}e(x)) \\ &= dF(x)i_{p(x)}e(x) + F(x) \wedge \left( \nabla i_{p(x)}e(x) + i_{p(x)}\nabla e(x) \right) \\ &= dF(x)i_{p(x)}e(x) + F(x) \wedge \left( d\ln\sqrt{|p(x)|}i_{p(x)}e(x) + i^2_{p(x)}A(x)e(x) \right) \\ &= dF(x)i_{p(x)}e(x) + F(x) \wedge \left( d\ln\sqrt{|p(x)|}i_{p(x)}e(x) + p(x)A(x)e(x) \right). \end{aligned} \quad (3.56)$$

The second term of $d_\nabla Re = 0$ reads

$$\begin{aligned} &F(x) \wedge \left( d\ln\sqrt{|p(x)|}i_{p(x)}e(x) + p(x)A(x)e(x) \right) \\ &= \left( dA(x) - A(x) \wedge d\ln\sqrt{|p(x)|} \right) \wedge \left( d\ln\sqrt{|p(x)|}i_{p(x)}e(x) + p(x)A(x)e(x) \right) \\ &= dA(x) \wedge d\ln\sqrt{|p(x)|}i_{p(x)}e(x) + dA(x) \wedge A(x)p(x)e(x) \\ &= dA(x) \wedge d\ln\sqrt{|p(x)|}i_{p(x)}e(x), \end{aligned} \quad (3.57)$$

where we used $d(A \wedge A) = dA \wedge A - A \wedge dA = 2dA \wedge A = 0$ since $A \wedge A = 0$. Therefore, we obtain the Bianchi identity that vanishes as follows

$$d_\nabla(Re) = \left( dF(x) + dA(x) \wedge d\ln\sqrt{|p(x)|} \right) i_{p(x)}e(x) = d^2 A(x) i_{p(x)} e(x) = 0 \quad (3.58)$$

due to $dF(x) = d^2 A(x) - dA(x) \wedge d\ln\sqrt{|p(x)|}$ from the definition of the field strength two-form $F(x)$, and de Rham cohomology for the exterior derivative $d^2 = 0$.

**Invariant action coupled to matter.** Consider a hypercomplex-valued scalar field of matter

$$\phi \equiv \frac{1}{\sqrt{2}} \rho(x) e^{i_{-f(x)} \sigma(x)} \in \mathbb{C}_{-f(x)}, \quad (3.59)$$

where $\rho, \sigma \in \mathbb{R}$. Assume that the field is charged under the relevant gauge group $G(M)$ (3.37) introduced in the previous section, so that

$$\phi(x) \longrightarrow \phi' = e^{-qi_{-f(x)}\theta(x)}\phi(x) \implies \rho'(x) = \rho(x), \quad \sigma'(x) = \sigma(x) - q\theta(x), \quad (3.60)$$

where $q$ is a charge of the field $\sigma$ which non-linearly transforms under the gauge group, and it should be noticed that the radial field $\rho$ is gauge invariant.

If we consider $f(x)$ as another scalar field degree of freedom, then we may have the following action

$$S = \int d^4x \left( D_\mu \phi^\star D_\mu \phi - \frac{f}{4g^2} F_{\mu\nu} F^{\mu\nu} + \frac{M^2}{2} \partial_\mu f \partial^\mu f - V(\phi^\star \phi, f) \right), \quad (3.61)$$

where $\star$ is the proper hypercomplex conjugate, $F_{\mu\nu}$ is the field strength tensor given in Eq. (3.51), $M$ is a mass scale of the gauge theory (mass dimension of $f$ is zero), and

$$\begin{aligned}
D_\mu \phi &= \frac{1}{\sqrt{2}} \left[ \nabla_\mu (\rho e^{i_{-f(x)}\sigma}) + i_{-f(x)} q A_\mu \rho e^{i_{-f(x)}\sigma} \right] \\
&= \frac{1}{\sqrt{2}} \left[ (\partial_\mu \rho) e^{i_{-f(x)}\sigma} + \rho (\nabla_\mu i_{-f(x)} \cdot \sigma + i_{-f(x)} \partial_\mu \sigma) e^{i_{-f(x)}\sigma} + i_{-f(x)} q A_\mu \rho e^{i_{-f(x)}\sigma} \right] \\
&= \frac{1}{\sqrt{2}} e^{i_{-f(x)}\sigma} \left( \partial_\mu \rho + i_{-f(x)} \rho (\sigma \partial_\mu \ln \sqrt{f} + \partial_\mu \sigma + q A_\mu) \right).
\end{aligned} \quad (3.62)$$

The total lagrangian is given by

$$\begin{aligned}
\mathcal{L} = &\frac{1}{2} |\partial_\mu \rho|^2 + \frac{f}{2} \rho^2 |\sigma \partial_\mu \ln \sqrt{f} + \partial_\mu \sigma + q A_\mu|^2 + \frac{M^2}{2} \partial_\mu f \partial^\mu f - V(\rho, f) \\
&- \frac{f}{4g^2} \left| \partial_\mu A_\nu - \partial_\nu A_\mu - (A_\mu \partial_\nu \ln \sqrt{f} - A_\nu \partial_\mu \ln \sqrt{f}) \right|^2.
\end{aligned} \quad (3.63)$$

If we take the following redefinitions

$$i_{-f} = \sqrt{f} \hat{i}_{-f}, \quad \theta = \frac{\hat{\theta}}{\sqrt{f}}, \quad A_\mu = \frac{\hat{A}_\mu}{\sqrt{f}}, \quad \sigma = \frac{\hat{\sigma}}{\sqrt{f}}, \quad f = \frac{\varphi}{M} \quad (3.64)$$

for $f > 0$, then these produce

$$g(x) = e^{i_{-f}\theta} = e^{\hat{i}_{-f}\hat{\theta}}, \quad \phi = \rho e^{i_{-f}\sigma} = \rho e^{\hat{i}_{-f}\hat{\sigma}}, \quad \hat{A}' = \hat{A} + \partial_\mu \hat{\theta}, \quad \hat{\sigma}' = \hat{\sigma} - q\hat{\theta} \quad (3.65)$$

and the gauge-invariant action with respect to the group $G(M)$ defined in Eq. (3.37):

$$\mathcal{L} = \frac{1}{2} |\partial_\mu \rho|^2 + \frac{1}{2} \rho^2 |\partial_\mu \hat{\sigma} + q \hat{A}_\mu|^2 - \frac{1}{4g^2} |\partial_\mu \hat{A}_\nu - \partial_\nu \hat{A}_\mu|^2 + \frac{1}{2} \partial_\mu \varphi \partial^\mu \varphi - V(\rho, \varphi/M), \quad (3.66)$$

which looks very similar to a gauged $U(1)$ action with one scalar degree of freedom. It is also easy to see that the number of field degrees of freedom is preserved during such redefinitions. Moreover, it is possible to compute the conserved Noether charge and current as usual in this gauge theory. Of course, the scalar and vector fields must have the form of Eq. (3.34) for them to be propagating with their plane-wave solutions in the framework of the fiber bundle of hypercomplex algebras. Last but not least, the scalar field $\varphi$ fully characterizes the whole of the physical action (3.66) since without the presence of $\varphi$, the other terms cannot be established. In addition, as a possible future investigation, it would be interesting to study another interpretation of the function $f(x)$ in this gauge theory, apart from the interpretation that $f(x)$ is given by another scalar field degree of freedom already discussed in this work.

## 4. Scalar field dynamics: solution to the vacuum selection problem

In this section, we discuss scalar field dynamics of the gauge theory built on the fiber bundle whose fibers consist of hypercomplex algebras. First, we have seen that once a particular function $f(x)$ is chosen, we are able to define a corresponding local hypercomplex imaginary unit $i_{p=-f(x)}$ and its spanning space $\mathbb{C}_{p=-f(x)}$, and then weave a set of the hypercomplex-algebra fibers over the base manifold into a fiber bundle. On this bundle, we define a hypercomplex-valued section or scalar field $\phi = \rho e^{i-f\sigma} = \rho e^{\hat{i}-f\hat{\sigma}}$. If the scalar field $\phi$ transforms under the gauge group $G(M)$, then the relevant gauge field $A_\mu$ (or $\hat{A}_\mu$ in canonically normalized form) corresponding to the group should be considered. In particular, it should be noticed that two different functions $f(x) \neq h(x)$ define two independent hypercomplex imaginary units $i_{-f(x)} \neq i_{-h(x)}$. That is, the function $f(x)$ completely characterizes the hypercomplex-valued scalar $\phi$ and gauge fields $A_\mu$, as well as its possible gauge group $G(M)$ (3.37). I shall call a codimension-1 hypersurface of the codomain of the function $f(x)$ over the base manifold $M$ as *F(unction-space)-brane* throughout this paper for later convenience. In addition, I call the F-brane with a constant value $f(x) = const.$ as *trivial F-brane*, while the F-brane with fluctuating values of the function $f(x)$ as *nontrivial F-brane*.

**Transmutation from nontrivial to standard F-brane.** We may classify gauge theories of conventional U(1) gauge group, like electromagnetism, in the mentioned terminology. Since a field charged under such U(1) is given by a local section of the complex U(1)-line bundle over the complex algebra $\mathbb{C}$ equipped with the imaginary unit $i$ such that $i^2 = -1$, it corresponds to the case when $f(x) = 1$. Hence, we may say that such U(1) theories are alive on a trivial F-brane of $f(x) = 1$. I shall call this special trivial F-brane as *standard F-brane* since most of the conventional gauge theories (related to the standard model) are built over the complex algebra. To be specific, in the gauge theory built on the fiber bundle of hypercomplex algebras, we have a crucial condition (3.32) that

$$f(x) = \varphi(x)/M > 0, \tag{4.1}$$

where $M$ is a positive constant as a mass scale of the theory. The authority of the condition $p = -f(x) < 0$ is to ensure that the kinetic terms of the relevant field degrees of freedom have right sign, i.e. to be ghost-free, which is possible only with $p = -f(x) < 0$. Keeping this in our mind, let us consider the Lagrangian (3.66) and the Higgs-like scalar potential shown in Fig. 2

$$V(\rho, f = \varphi/M) = \alpha\left(1 - (\varphi/M)^2\right)^2 + V(\rho), \tag{4.2}$$

where $\alpha$ is a constant. This potential has two degenerate vacua at $\varphi = \pm M$. We now suppose that the dynamical field $f(x)$ rolls down to one of the possible ground states with minimal energy. Then, at the true minimum, the function $f(x)$ gets frozen at a constant value of $f(x)$ after rolling down. This leads to the situation that a nontrivial F-brane with an arbitrary $f(x)$ transmutes into a trivial F-brane with $f(x) = 1$ or equivalently $\varphi = +M$ (called standard F-brane), which is depicted in Fig. 1. I shall call this phenomenon *transmutation of F-brane*.

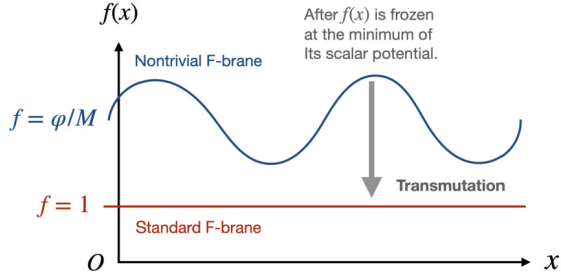

Fig. 1. Transmutation from nontrivial to standard F-brane.

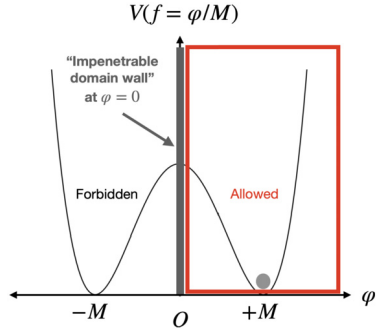

Fig. 2. The impenetrable domain wall prohibits phase transition between two degenerate vacua.

**Vacuum selection with priority and impenetrable domain wall caused by intrinsic even parity under a $\mathbb{Z}_2$ symmetry.** As a key property of the gauge theory built on the fiber bundle, I present how one particular vacuum can be selected with priority between two degenerate physical vacua through an intrinsic even parity of the scalar field $f(x)$ under a $\mathbb{Z}_2$ symmetry. At first glance, it seems plausible for the (dimensionless) field $f(x)$ to allow its even and odd parities under a $\mathbb{Z}_2$ symmetry once its scalar potential is given by the scalar potential like Eq. (4.2) which is $\mathbb{Z}_2$-symmetric under the sign change of the field, $f \to -f$. However, in fact, *any $\mathbb{Z}_2$ symmetry is intrinsically forbidden in this gauge theory if the field $f(x)$ has odd parity, regardless of the $\mathbb{Z}_2$ invariance of its scalar potential*. This is because from the beginning we have defined the hypercomplex imaginary unit $i_p$ to be characterized by the condition (3.32) for some positive-definite function $f(x)$. In other words, such transformation $f \to -f$ leads to the worst situation that some relevant field degrees of freedom change into ghosts. In this respect, the field $f(x)$ must be even under a $\mathbb{Z}_2$ symmetry, i.e. $f \to f$, which is supported by the condition (3.32). Also, it turns out that there exists an impenetrable domain wall at the point $f = 0$, i.e. $\varphi = 0$ for the scalar potential as shown in Fig. 2 in that it is impossible for phase transition between the two degenerate vacua to take place. Conclusively, in this gauge theory, the field $f(x)$ can be permanently frozen without tunneling across the domain wall only at its positive value in which the minimum of the scalar potential is located.

**Improved spontaneous-symmetry-breaking scenario with no vacuum selection problem.** During the transmutation, a gauge group $G(M)$ in Eq. (3.37) with respect to $f(x)$ also transmutes into a usual gauge group $U(1)$ over the complex algebra $\mathbb{C}$ with the constant $f(x) = 1$, i.e.

$$G(M)|_{p=-f(x)} \longrightarrow G(M)|_{p=-f(x)=-1} = U(1)|_{\text{on standard F-brane}} \neq U(1)|_{\text{normal}}. \quad (4.3)$$

As for this, I would like to point out that the conventional U(1) gauge theory on the complex line bundle (i.e. $f(x) = 1$ living on the standard F-brane from the beginning) must be distinct from the gauge theory of a non-constant $f(x)$ with $G(M)$ (3.37) living on nontrivial F-brane that may transmute into the standard F-brane after $f(x)$ is dynamically frozen at its vacuum where $f(x) = 1$ is set. I emphasize that the two cases are completely independent in that the latter additionally entails the condition (3.32), while the former does not. Again, this feature is a big difference between the newly-constructed here and normal gauge theories. Respecting the above argument, the action (3.66) can become such usual U(1) action (but entailing the condition (3.32)) only if the scalar field $\varphi$ is frozen at its vacuum expectation value such that $\varphi = M$ since this transmutes hypercomplex algebra $\mathbb{C}_{p=-f(x)}$ into the complex one $\mathbb{C}_{p=-f(x)=-1}$ with $i = \sqrt{-1}$ such that $i^2 = -1$. (In this situation, the field $\varphi$ may be a component of some field multiplet which is charged under another gauge group to be spontaneously broken later.) The resulting action is still gauge-invariant under the U(1) group left after the transmutation. On the other hand, as another possibility, if we assume that $f(x) = \rho(x)$, then the corresponding Lagrangian is given by

$$\mathcal{L} = \frac{1}{2}|\partial_\mu \rho|^2 + \frac{1}{2}\rho^2 |\partial_\mu \hat{\sigma} + q \hat{A}_\mu|^2 - \frac{1}{4g^2}|\partial_\mu \hat{A}_\nu - \partial_\nu \hat{A}_\mu|^2 - V(\rho), \quad (4.4)$$

whose field components are specified by the hypercomplex-valued scalar field $\phi = \rho e^{\hat{i} - \rho(x)\hat{\sigma}}$. As usual, if we consider Higgs-like scalar potential $V(\rho) = \alpha(\rho^2 - v^2)^2$ for some constants $\alpha, v$, then we are able to perform spontaneous symmetry breaking (SSB) with respect to the gauge group $G(M)$ (3.37), but selecting one special vacuum thanks to the condition (3.32). In this sense, the gauge theory on the fiber bundle of hypercomplex algebras can be considered better than the existing gauge theory with SSB which has the vacuum selection problem. Lastly, I hope that this framework studied here would be used as alternative of the conventional gauge theory with SSB, and that more physical implications of this framework would be further investigated in many aspects not treated here.

**CRediT authorship contribution statement**

**Hun Jang:** Conceptualization, Formal analysis, Investigation, Methodology, Visualization, Writing – original draft, Writing – review & editing.

**Declaration of competing interest**

The authors declare that they have no known competing financial interests or personal relationships that could have appeared to influence the work reported in this paper.

**Data availability**

No data was used for the research described in the article.

**Acknowledgements**

H.J. was supported by a James Arthur Graduate Associate (JAGA) Fellowship from Center for Cosmology and Particle Physics in Department of Physics at New York University at the beginning of this work, and also supported by a Riemann Fellowship from Riemann Center for Geometry and Physics at Leibniz University Hannover. H.J. would like to thank the Institute for Theoretical Physics at the Riemann Center for its kind hospitality at the completion of this work.